\documentclass[twocolumn]{svjour2}

\usepackage{graphicx}
\journalname{Plasmonics}

\begin{document}

\title{Plasmonic engineering of metal nanoparticles for enhanced fluorescence and Raman scattering}

\author{N.\ I.\ Cade \and
        T.\ Ritman-Meer \and
        D.\ Richards}
\institute{N.\ I.\ Cade \at
        Department of Physics, King's College London, Strand, London WC2R 2LS, UK\\
        Tel.: +44 207848 1531\\
        Fax: +44 207848 2420\\
\email{nicholas.cade@kcl.ac.uk}
}
\date{Received: date / Accepted: date}

\maketitle

      \begin{abstract}

We have investigated the effects of tuning the localized surface plasmon resonances (LSPRs) of silver nanoparticles on the fluorescence intensity, lifetime, and Raman signal from nearby fluorophores. The presence of a metallic structure can alter the optical properties of a molecule by increasing the excitation field, and by modifying radiative and non-radiative decay mechanisms. By careful choice of experimental parameters we have been able to decouple these effects. We observe a four-fold increase in fluorescence enhancement and an almost 30-fold increase in decay rate from arrays of Ag nanoparticles, when the LSPR is tuned to the emission wavelength of a locally situated fluorophore. This is consistent with a  greatly increased efficiency for energy transfer from fluorescence to surface plasmons. Additionally, surface enhanced Raman scattering (SERS) measurements
show a maximum enhancement occurs when both the incident laser light and the Raman signal are near resonance with the plasmon energy. Spatial mapping of the SERS signal from a nanoparticle array reveals highly localized differences in the excitation field resulting from small differences in the LSPR energy.

\keywords{Fluorescence enhancement, Fluorescence lifetime, LSPR, Nanoparticles, SERS}
\end{abstract}

\section{Introduction}
\label{sec:intro}

Noble metal nanoparticles are being used increasingly in spectroscopic techniques with biomedical applications, including detection, labelling, cell tagging and sorting, imaging enhancers, and as therapeutic agents \cite{Stuart2005}. When metal nanoparticles are excited by electromagnetic radiation of a certain wavelength, their conduction electrons exhibit collective oscillations known as a localized surface plasmon resonance (LSPR). For a specific nanoparticle, the wavelength of the LSPR is sensitive to a variety of factors, such as the size, shape, interparticle spacing, and dielectric environment \cite{vanduyne01,mock02,schatz03,Jin2003,Rechberger2003,Murray2006,Stranik2007,Gonzalez2007}. The presence of a plasmonic structure can dramatically alter the optical properties of a locally situated molecule \cite{barnes98,fort08}: The excitation field is strongly enhanced due to both the surface plasmons and the lightning rod effect for highly curved metal surfaces. This has important consequences in Raman spectroscopy, as the signal can be enhanced by many orders of magnitude via surface enhanced Raman scattering (SERS) from metal nanoparticles \cite{nie97}. Furthermore, metallic structures can also  modify  both the radiative and the nonradiative decay rates for a molecule \cite{lakowicz02,dulkeith02,kuhn06,gerber07,Muskens2007}, resulting in enhanced fluorescence emission and greater photostability by reducing the excited state lifetime.

Recent improvements in nanofabrication methods have enabled tuning of the LSPR to optimize the optical enhancement properties, for a wide range of nanostructured systems; these include nanoshells \cite{Jackson2004}, nanocrescents \cite{Kim2006}, nanovoids \cite{perney06}, and various other types of nanoparticles \cite{Felidj2003,Corrigan2005}. Here, we investigate the effects of tuning the plasmon resonance  on fluorescence intensity and lifetime, and Raman scattering, for two types of nanostructures. By careful choice of experimental parameters we have been able to separately investigate modifications  arising from enhancement of the excitation field and from modified radiative decay mechanisms. For fluorophores on nanostructures with an LSPR matching the fluorescence emission, we have observed a four-fold increase in intensity and an almost 30-fold increase in decay rate relative to unmodified molecules. SERS measurements indicate that enhancement of the excitation field is also highly sensitive to the overlap between the LSPR and the laser / Raman line.


\section{Experimental Details}
\label{Sec:experiment}

\begin{figure}[tb]
\begin{center}
\includegraphics[width=8cm]{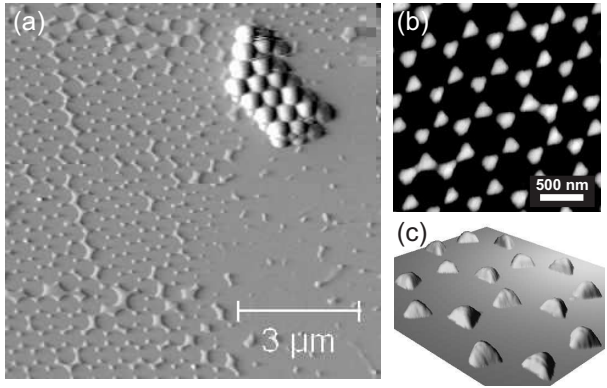}
\end{center}
\caption{(a) AFM image of a nanoparticle array created by a single layer of latex spheres. Some residual spheres still remain in the top right corner. (b) Zoom image from a region on the left side of (a).  (c)  3-dimensional representations of (b). } \label{afm}
\end{figure}

Glass coverslips were immersed in a solution of 3:1 sulphuric acid and hydrogen peroxide for an hour. After  rinsing in deionized water, the coverslips were transferred to a solution of ammonium hydroxide, hydrogen peroxide, and water in a 1:1:4 ratio, and heated at 80~$^{\circ}$C for an hour. The coverslips were then thoroughly rinsed in deionized water again. This process removes organic matter and hydroxylates the glass surface. 500~nm diameter latex spheres (Invitrogen) were diluted in deionized water and drop-cast onto a coverslip, then left to dry; as the water evaporated the spheres assembled into a  close-packed lattice \cite{vanduyne01}. A 0.5~nm layer of chromium was then deposited onto the slide by thermal
evaporation in a vacuum chamber; this increases the adhesion of silver which was subsequently deposited to a thickness of 40~nm. The latex
spheres were removed by sonication in chloroform, leaving a continuous periodic array of Ag nanoparticles. All chemicals mentioned in this report were purchased from Sigma Aldrich unless explicitly stated.

Figure \ref{afm}(a) shows an atomic force microscopy (AFM) image of a boundary region of nanoparticles, formed from a single layer of latex spheres (left side). There is a small region of residual latex spheres in the top right corner.
Figure \ref{afm}(b) shows individual particles within the array, which have a base length of $\sim$120 nm; Figure \ref{afm}(c) is a 3-dimensional representation showing the formation of well defined facets.

To investigate the plasmonic effects of the nanoparticles on nearby fluorophores, two identical samples were fabricated as described above. One sample was kept as a control, and the second was modified by growing a thin capping layer of SiO$_2$: 200~$\mu$l of undiluted (3-amino\-propyl)\-tri\-methoxy\-silane (APS) was spin-coated onto the sample; this has a strong chemical affinity for metal surfaces and acts as a primer to render the surface vitreophilic. The sample was then left overnight in a solution of sodium silicate and ammonium hydroxide at pH~11.75, which produced a  SiO$_2$ shell a few nanometers thick \cite{Liz-Marzan1996}.  Extinction spectroscopy (Perkin-Elmer Lambda 800) was used to determine the LSPRs of the samples.

For fluorescence measurements, a chlorophyll derivative pheophorbide-a (PPa) was dissolved to a concentration of $5 \times 10^{-5}$ M in a 0.75~$\%$ solution of poly\-(methyl\-meth\-acrylate) (PMMA) in toluene. 200~$\mu$l of this solution was spin-coated on both of the different nanoparticle arrays at 2000~rpm for 1 minute; this produced a thin film ($\sim$20~nm) of PPa in a PMMA matrix, uniform across the sample. Each sample was prepared for high resolution optical measurements by applying a thin layer of index-matching polymer mountant (Mowiol 4-88) to a clean coverslip and affixing this on top of the nanoparticle array / PPa film.

Fluorescence lifetime and intensity images were obtained using a scanning confocal microscope  (Leica TCS SP2, 63$\times$ water objective,
1.2 NA), with a synchronous time correlated single photon counting module (Becker and Hickl SPC-830). Excitation was with a 488~nm
continuous-wave (CW) Ar$^{+}$ laser (intensity) or 467~nm 20~MHz pulsed diode laser (Hamamatsu PLP) (lifetime), at a power far below fluorescence saturation in both cases. Decay transients were analyzed using iterative reconvolution (TRI2 software), including an instrumental response of $\sim$90~ps.

For Raman investigations, an array of Ag nanoparticles was created using 650~nm latex spheres, as described above. Rhodamine~6G (R6G) dye was then deposited onto the array by vacuum sublimation to produce a  uniform monolayer coverage. Spatially averaged Raman spectra were acquired using a Renishaw spectrometer with a 20$\times$ objective lens (Leica, 0.4 NA). To reduce fluorescence background, excitation was at 752~nm using a home-made tunable CW Ti:sa laser. To obtain high resolution SERS maps, a 100$\times$ objective lens (Leica, 1.4 NA) was used with a high precision scanning stage (Prior H101).

\section{Fluorescence Enhancement}
\label{sec:fluo}

\begin{figure}[tb]
\begin{center}
\includegraphics[width=7.5cm]{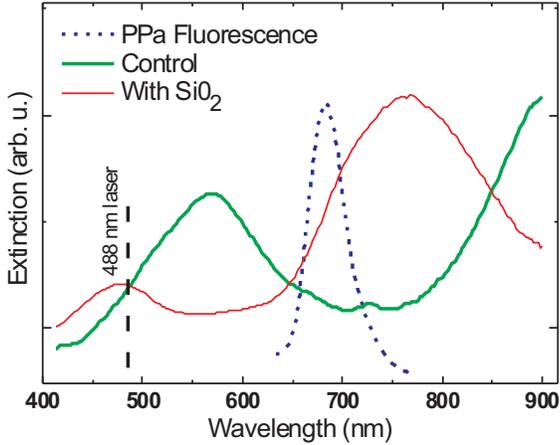}
\end{center}
\caption{ Ensemble extinction spectra from nanoparticle arrays with and without SiO$_2$. The dotted line is the fluorescence emission spectrum from PPa. The vertical dashed line is the 488 nm CW laser.} \label{extinction}
\end{figure}

Extinction spectra from the two nanoparticle samples are shown in Figure \ref{extinction}. Each spectrum has two main extinction resonances corresponding to dipole ($>$650~nm) and quadrupole ($<$650~nm) LSPR modes \cite{schatz03,nelayah07}. The addition of the SiO$_2$ capping layer has produced a large shift ($\sim$150~nm) in the position of both LSPR modes; crucially, the main dipole peak has a much larger overlap with the emission spectrum of PPa, shown as a dotted line in Figure \ref{extinction}.  AFM analysis of the nanoparticles after SiO$_2$ growth shows a negligible change in height, but a more spheroidal shape than the control nanoparticles. Extinction measurements taken after the addition of the PPa film and mountant showed a negligible shift and slight narrowing of the LSPR peak, for both samples. This suggests that, for the thin SiO$_2$ layer used here, the shift in LSPR position due to a change in the dielectric environment is much less than the shift resulting from other factors such as size and shape \cite{vanduyne01,mock02}.

Importantly, the laser excitation wavelengths are far from the dipole LSPR peaks shown in Figure \ref{extinction}, in a region where the extinction characteristics are almost identical for both samples. The excitation is also off-resonance with all PPa absorption peaks \cite{Korth1998}. Any effects originating from an increased excitation field should be greatly reduced, and similar for both samples \cite{kuhn06}; hence, any differences in the fluorescent properties of the two samples will result from changes in resonant coupling of emission radiation to plasmons in the metal nanoparticles, as discussed below.

\begin{figure}[tb]
\begin{center}
\includegraphics[width=8cm]{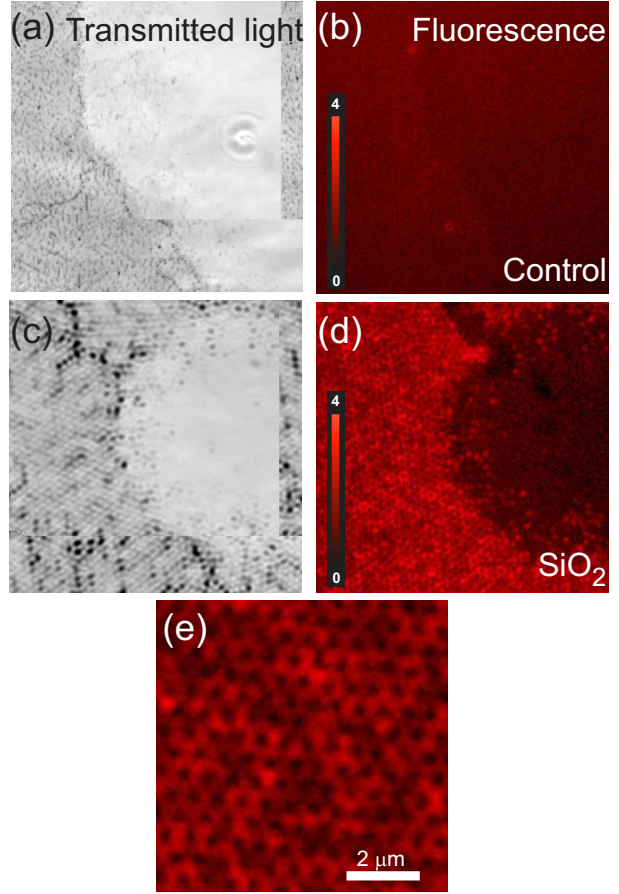}
\end{center}
\caption{ (a) Transmitted laser light image of a boundary region between single layer nanoparticles and glass, for the control sample. (b)
Corresponding confocal  fluorescence intensity map of PPa, for the same region. (c) and (d); as (a) and (b) for the SiO$_2$ sample. The intensity scale in (b) and (d) has been normalized to the signal from the glass. (e)~Zoom of nanoparticle region in (d); \emph{cf}.\ Figure \ref{afm}(b).} \label{fluorescence}
\end{figure}

Figure \ref{fluorescence}(a) is a transmitted light image of a boundary region between  nanoparticles and glass similar to that of Figure \ref{afm}(a), for the control sample. The corresponding confocal PPa fluorescence signal detected from this region is shown in Figure \ref{fluorescence}(b). Figures \ref{fluorescence}(c) and (d) are transmitted and fluorescence images for the SiO$_2$ capped nanoparticles, respectively. Figure \ref{fluorescence}(e) is a high resolution image from a region in (d). Spectral analysis has verified that the emission originates from PPa and is not caused by photoluminescence from the nanoparticles or scattered laser light. In contrast to the control sample, the fluorescence images show highly localized enhancement from fluorophores close to nanoparticles; the intensity scale has been normalized between Figures \ref{fluorescence}(b) and (d) to the signal from the glass. These images show a four-fold increase in emission intensity for fluorophores on nanoparticles modified by SiO$_2$ relative to the control sample.  The actual enhancement in the vicinity of the nanoparticles is much greater than that measured in the far-field: within a confocal  spot of $\sim$400 nm diameter, only those fluorophores within approximately 10--20~nm of the nanoparticle will show significant enhancement \cite{anger06,rm07}; this corresponds to $<$5~$\%$ of the excitation region.  For the control sample, the essentially uniform fluorescence intensity indicates that any enhancement effects from the nanoparticles are negated by quenching close to the metal. Further tests were done to verify that this increase in enhancement is a direct result of the LSPR shift, as described in the next section.

\section{Fluorescence Lifetime Modification}
\label{sec:lifetime}

Using time-resolved fluorescence measurements, it is possible to further verify the separate contributions to the fluorescence enhancement arising from modified excitation and decay channels in the presence of a metal nanoparticle.

\begin{figure}[tb]
\begin{center}
\includegraphics[width=8cm]{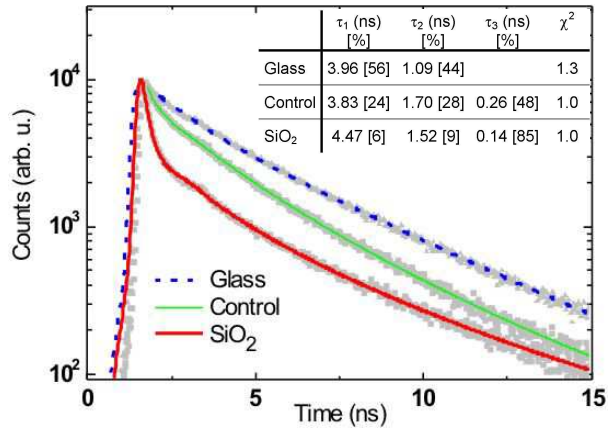}
\end{center}
\caption{Fluorescence lifetime decay transients from PPa on glass, and on nanoparticles with and without SiO$_2$. Curves have been normalized to the same initial amplitude. Inset: constituent lifetime components and their relative amplitudes (square brackets), obtained from fits to the transients.} \label{lifetime}
\end{figure}

Figure \ref{lifetime} shows spatially integrated decay transients from the PPa film on glass and on nanoparticles with and without SiO$_2$. Each transient has been fitted with a bi- or tri-exponential decay: the corresponding lifetime components, their respective relative amplitudes, and reduced $\chi^{2}$ values are summarized in the table. The errors in the lifetimes are all $\leq\pm0.04$~ns.

The fluorescence decay of PPa on glass has a biexponential nature with 4~ns and 1~ns components of approximately equal amplitude; this is due to concentration quenching and thin film effects. Lifetime measurements were taken of PPa in 0.75~$\%$ PMMA in toluene, before spin-coating, at concentrations of $5 \times 10^{-5}$~M and $1 \times 10^{-6}$~M. These had mono-exponential decays with lifetimes of 5.6~ns and 6.0~ns, respectively, which are consistent with previously reported values for PPa in different environments and concentrations \cite{Korth1998,Roder2000}. The quantum yield of PPa is also significantly reduced in a thin film due to interactions with the matrix and other dye molecules: Lagorio \textit{et al}.\ have reported values of around 0.05 for PPa samples exhibiting similar lifetimes to those measured here for PPa on glass \cite{Lagorio2001}.

For PPa on nanoparticles without SiO$_2$, the fluorescence decay has two components similar to those on glass, and an additional short lifetime component of 260~ps which accounts for approximately half of the total amplitude. However, the decay transient for PPa on the SiO$_2$ sample is significantly different, and essentially comprises just a single lifetime of 140~ps. This indicates a $\sim$30-fold increase in decay rate relative to unmodified fluorophores on glass. These results are consistent with spatially resolved lifetime measurements we have taken for R6G dye on nanoparticles: we observed an order of magnitude increase in decay rate from a localized volume around the nanoparticles due to the additional metal-induced radiative mechanisms \cite{rm07}.

Lakowicz  has proposed that excited fluorophores can nonradiatively transfer energy to surface plasmons, which in turn radiate the emission of the coupling fluorophores \cite{Lakowicz2005}. As the transfer efficiency approaches 100~$\%$, the effective quantum yield of the fluorophore approaches unity \cite{Aslan2005}. For the samples investigated here, the increased decay rate and intensity for the SiO$_2$ sample is consistent with a more efficient coupling of the PPa fluorescence to surface plasmons, leading to an order of magnitude increase in quantum yield.

As the SiO$_2$ capping layer is relatively thin ($<$5~nm), fluorescence quenching effects will be significant for both nanoparticle arrays \cite{Stranik2007}; however, quenched fluorophores will give a negligible contribution to the total fluorescence lifetime signal and the effects can be disregarded. The residual longer lifetime components seen for the  SiO$_2$ sample are probably from unmodified fluorophores in the film far from the nanoparticles \cite{Muskens2007}, as discussed in the previous section.

Further tests were done to check that the observed modifications to fluorescence intensity and lifetime originated from the shift in plasmon resonance between the samples, and were not caused by the SiO$_2$ itself: Additional nanoparticle arrays were fabricated with varying amounts of Ag deposited but no SiO$_2$. After measuring the extinction spectra, a sample was chosen with an almost identical LSPR peak to that of the SiO$_2$ sample. A thin film of PPa in PMMA was then deposited, and fluorescence intensity and lifetime measurements were taken as before. The results are very similar to those shown in Figures \ref{fluorescence} and \ref{lifetime} for the SiO$_2$ sample: strong fluorescence enhancement is observed around the nanoparticles and the fluorescence decay is essentially monoexponential with a lifetime of 150~ps.

\section{Surfaced enhanced Raman scattering}
\label{sec:sers}

\subsection{Nanoparticle arrays}

In contrast to the mechanisms discussed in the last section, the enhanced Raman signal seen in SERS measurements is due to the
increase in  intensity of the local excitation field around an individual nanoparticle: maximum SERS enhancement should occur when the LSPR  coincides with a wavelength between that of the exciting laser and the Raman line \cite{Felidj2003}. With these optimum excitation conditions, the  SERS enhancement depends on the fourth power of the total electric field $E$, due to an increase in both the incident laser field and the emission field at the Raman frequency \cite{kneipp02}.

\begin{figure*}[tb]
\begin{center}
\includegraphics[width=17.5cm]{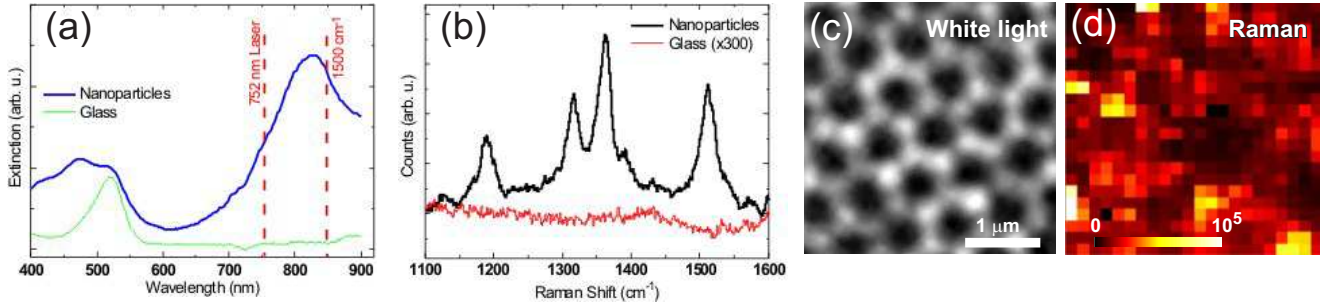}
\end{center}
\caption{ (a) Extinction spectra from R6G dye on a nanoparticle array and on glass. The vertical dashed lines show the  wavelength of the laser and corresponding 1500~cm$^{-1}$ shift. (b) Raman spectra from R6G on nanoparticles and glass. (c)~Reflected white-light image of the nanoparticle array. (d) Intensity map of the 1500~cm$^{-1}$ Raman line, from a region similar to (c).} \label{raman}
\end{figure*}

To investigate the suitability of these nanoparticles for use in SERS-based biosensors, an array of Ag nanoparticles was covered with a thin film of R6G, as described previously. Figure \ref{raman}(a) shows ensemble extinction spectra from the R6G covered nanoparticles  and from R6G on glass. Figure \ref{raman}(b) shows the Raman spectra from R6G on a region of nanoparticles (2~s acquisition time) and from R6G on glass (10~min acquisition time). As no Raman signal was obtained from the glass region, it is not possible to calculate an absolute value for the enhancement factor; however, a lower limit of $10^5$ can be estimated using the noise level in the glass spectrum. To further investigate the SERS enhancement, high resolution spatial maps were obtained of the Raman signal over the nanoparticles; Figure \ref{raman}(d) shows the integrated intensity under the 1500 cm$^{-1}$ Raman line, for a region of nanoparticles similar to \ref{raman}(c).

Figure \ref{raman}(a) shows that the position of the ensemble LSPR peak for this nanoparticle array lies between the exciting laser and the Raman lines; hence we would expect a very strong overall SERS enhancement, as seen in Figure \ref{raman}(b). However, the LSPR of an \emph{individual} nanoparticle is highly sensitive to small changes in its size, shape, and local environment \cite{Sherry2006}; hence, variations between nanoparticles will lead to highly localized differences in the field enhancement. This is particularly noticeable in SERS measurements, due to the $E^4$ dependence, and results in the creation of optical ``hot spots", as seen in Figure \ref{raman}(d) \cite{gresillon99}.

\subsection{Thin Ag films}

\begin{figure*}[tb]
\begin{center}
\includegraphics[width=16cm]{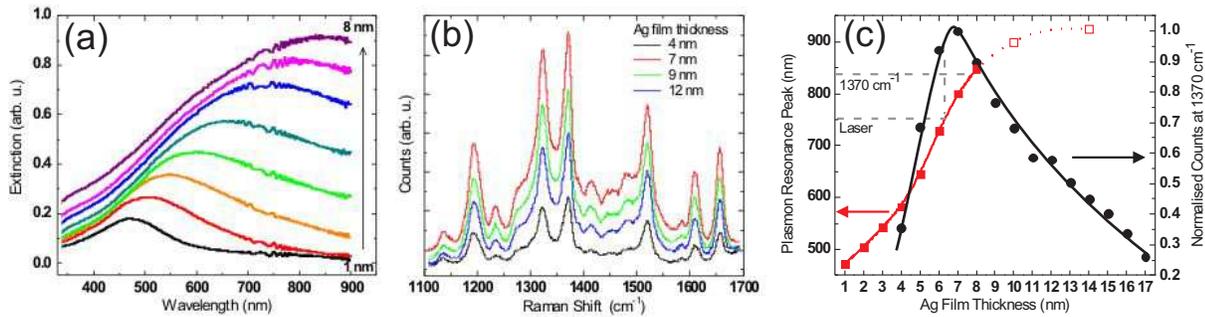}
\end{center}
\caption{ (a) Extinction spectra from Ag films with nominal thickness from 1~nm -- 8~nm. (b) SERS spectra from R6G on Ag films of various thickness. (c) Left axis: Plasmon resonance peak position vs film thickness, obtained from (a). The dotted region shows estimated positions. Right axis: Normalized intensity of the Raman peak at 1370~cm$^{-1}$ vs film thickness, obtained from (b). The solid lines are guides to the eye. The dashed lines show the  wavelength of the laser and corresponding 1370~cm$^{-1}$ shift.} \label{ag_films}
\end{figure*}

The effect of tuning the plasmon resonance on SERS enhancement was also investigated: a sample was prepared by thermal deposition of Ag films with a range of nominal thicknesses, on the same glass slide.  Extinction spectra were then taken, as shown in Figure \ref{ag_films}(a): the position of the plasmon resonance shifts from below 500~nm to above 900~nm with increasing film thickness.

A thin layer of R6G dye was deposited by vacuum sublimation, creating a uniform coverage over the different Ag films. Raman spectra were acquired using a 752~nm laser and a 20$\times$ objective, and are shown in Figure \ref{ag_films}(b) for a selection of Ag films. The data from Figures \ref{ag_films}(a) and (b) are summarized in \ref{ag_films}(c): The left axis (squares) shows the shift in the plasmon peak with increasing film thickness, with the dotted region indicating estimated positions outside of the instrument range. The right axis (circles) shows the intensity of the Raman peak at 1370~cm$^{-1}$, normalized to the maximum value.  The dashed lines indicate the wavelength of the laser and a corresponding shift of 1370~cm$^{-1}$.

With increasing film thickness, the  SERS enhancement shows a sharp maximum at 7~nm and decreases again for thicker films. As discussed above, when both the incident laser light and the Raman signal are near resonance with the plasmon energy, the SERS enhancement will have a maximum $E^4$ dependence; the data in Figure \ref{ag_films}(c) show very good agreement with this. A rigorous analysis of the data requires additional factors to be considered, such as grain size, morphology, and surface coverage; this will be reported elsewhere.


\section{Conclusion}

We have presented an investigation into the effects of tuning the plasmon resonance of metal nanoparticles on the optical properties of nearby fluorophores. Two essentially identical arrays of nanoparticles were prepared, and the LSPR position of one was shifted to produce a large overlap with the emission from a thin layer of the chlorophyll derivative PPa.  For the modified sample, we observe a four-fold increase in fluorescence intensity and an almost 30-fold increase in decay rate relative to unmodified fluorophores. By careful choice of experimental parameters we have been able to decouple the influence of changes in the excitation field; the results are consistent with a greatly increased efficiency for energy transfer from fluorescence to surface plasmons, for the optimized nanoparticles. The effect of tuning the LSPR on SERS signal was also investigated; in this case  a maximum enhancement occurs when both the incident laser light and the Raman signal are near resonance with the plasmon energy. Spatial mapping of the SERS signal from a nanoparticle array shows an enhancement of more than 5 orders of magnitude in small hot spots, due to highly localized differences in the excitation field. These results have important implications for a wide range of fluorescence-based technologies; in particular, plasmonic engineering of nanostructures will have a crucial role in the  development of novel biomedical applications, including biosensors, high throughput screening systems, and non-invasive therapy.

\section{Acknowledgements}

The authors would like to thank K.\ Suhling and M.\ Green (Dept.\ of Physics, King's College London) for helpful discussions and advice. This work was supported by the EPSRC (UK).

\end{document}